\documentclass[11pt]{article}
\usepackage{amssymb,amsmath}
\makeatletter
\@addtoreset{equation}{section}
\makeatother

\newtheorem{thm}{Theorem}

\newcommand{\auth}{Bartholomew Furrow$^\dagger$}
\newcommand{\epsinv}{\epsilon^{-1}}
\newcommand{\sle}{\sqrt{\lg\epsinv}}
\newcommand{\ggdef}{$\mathbb{G}=(\mathbb{V},\mathbb{E})$}
\newcommand{\length}{\text{length}}
\newcommand{\ith}{^\text{th}}
\newcommand{\vect}{\overrightarrow}
\newcommand{\modd}[1]{\left|d[#1]\right|}
\newcommand{\lgl}{\log_{\lambda}}

\newcommand{\bbht}{{\it BBHT}}
\newcommand{\bcwz}{{\it BCWZ}}
\newcommand{\minfind}{{\it minfind}}
\newcommand{\findall}{{\it findall}}
\newcommand{\mindiff}{{\it mindiff}}
\newcommand{\findsol}{{\it findsol}}
\newcommand{\bfs}{{\it BFS}}
\newcommand{\dfs}{{\it DFS}}

\newcommand{\coinchange}{{\it coinchange}}
\newcommand{\spnw}{{\it SPNW}}
\newcommand{\apsp}{{\it APSP}}

\newcommand{\maxpoints}{{\it maxpoints}}
\newcommand{\subarraysum}{{\it subarray-sum}}

\newlength{\cwidth}
\newcommand{\cents}{\settowidth{\cwidth}{c}%
\divide\cwidth by2
\advance\cwidth by-.1pt
c\kern-\cwidth
\vrule width .1pt depth.2ex height1.2ex
\kern\cwidth}

\begin{document}

\begin{flushright}

\end{flushright}

\vspace{10pt}

\begin{center}
{\large {\bf A Panoply of Quantum Algorithms }}

\vspace{20pt}
\auth

\vspace{10pt}

{\it $^\dagger$ Department of Physics and Astronomy, University of British Columbia,\\
Vancouver, British Columbia  V6T 1Z1, Canada}

{\it
   {\rm E-mail: \texttt{furrow@phas.ubc.ca}}
  }


\vspace{40pt}

\underline{ABSTRACT}

We create a variety of new quantum algorithms that use Grover's algorithm and similar techniques to give polynomial speedups over their classical counterparts.  We begin by introducing a set of tools that carefully minimize the impact of errors on running time; those tools provide us with speedups to already-published quantum algorithms, such as improving D\"urr, Heiligman, H\o yer and Mhalla's algorithm for single-source shortest paths\cite{dhhm} by a factor of $\lg N$.  The algorithms we construct from scratch have a range of speedups, from $O(E)\rightarrow O(\sqrt{VE\lg V})$ speedups in graph theory to an $O(N^3)\rightarrow O(N^2)$ speedup in dynamic programming.

\end{center}

\pagebreak

\section{Introduction}

This paper introduces several new quantum algorithms which are polynomially faster than their classical counterparts.  We introduce these through the use of Grover's algorithm and its descendants as introduced by Boyer, Brassard, H\o yer and Tapp (modified in Appendix \ref{bbht_error}) and Buhrman, Cleve, de Wolf and Zalka\cite{bbht,bcwz}.  We begin by introducing some basic tools, such as minimum-finding, that use Grover's directly; in the construction of those tools we pay particular attention to the probability with which they fail, and make their running time depend as little as possible on the desired probability of failure.

After introducing our tools we cast our gaze over several fields, striving to address a variety of classical algorithms, especially those that are illustrative of a particular problem type.  We find $O(\sqrt{E/V})$ improvements in some important graph theory algorithms, and also examine some already-published quantum algorithms in graph theory\cite{dhhm, as}, giving them logarithmic speedups by improving how they deal with errors.  After that we examine some algorithms in computational geometry and dynamic programming, where we find perhaps our most impressive individual results: $O(N)$ and $O(\sqrt{N})$ improvements over the best-known classical algorithms.  For completeness' sake, we include an appendix of comments and caveats (Appendix \ref{caveats}), which contains a section on some of the notation used here with which physicists might be unfamiliar.

For a summary of our algorithms' running times compared to those for classical solutions to the same problems, please see our conclusions in section \ref{conclusions}.

\section{Grover's algorithm}

We make extensive use of descendants of Grover's search algorithm\cite{grover}.  Grover's algorithm works as follows: we are given a binary function (one that returns only 0 or 1), $F,$ over a domain of size $N,$ with only one value for $x$ such that $F(x)=1$ (we will call such values ``solutions for $F$'').  Grover found that it took just $O(\sqrt{N})$ calls to $F$ to find a value $x$ such that $F(x)=1$.  To find such a value of $x$ classically, assuming no knowledge of the properties of $F,$ would take $O(N)$ calls to $F$.  Since its initial introduction by Grover, several improvements have been made to the algorithm; here we restate the results we will use, which we will refer to by the initials of their authors:

\begin{itemize}
\item {}{\bf BBHT:} If there are $M>0$ solutions to $F$ in the domain (we do not need to know $M$), the \bbht\cite{bbht} search algorithm returns one random such element after $O(\sqrt{N/M})$ calls to $F$.  There is probability $\approx .5M^{-.93}$ that it will fail, returning the special value {\it false} after $O(\sqrt{N})$ calls to $F$.  If $M=0,$ it returns {\it false} in $O(\sqrt{N})$ calls to $F$.  Note that in their original paper, Boyer, Brassard, H\o yer and Tapp do not discuss the probability of failure and the $M=0$ case in depth; we do so in Appendix \ref{bbht_error}.

\item {}{\bf BCWZ:} The \bcwz\cite{bcwz} search algorithm is passed a parameter $\epsinv$ and returns a random solution to $F$ after $O(\sqrt{N\lg{\epsinv}})$ calls, provided that such a solution exists.  There is a probability $\epsilon$ that it will fail, in which case it returns {\it false}.  If $M=0,$ it returns {\it false} in $O(\sqrt{N\lg\epsinv})$ calls to $F$.
\end{itemize}

\section{Algorithmic tools} \label{tools}

Here we present some basic algorithms, founded on the above primitives, that serve as subroutines to be used throughout this paper (where they will be referred to by their abbreviated names, found in the subsection headers).  We begin by noting that if an algorithm is to be run $R$ times, and we want it to succeed all $R$ times with some constant probability, the algorithm must have probability $\epsilon < 1/R$ of failure.  Because of this, we will sometimes talk about $\epsinv$ being polynomial, and we carefully formulate algorithms in this section to minimize the dependence of running time on $\epsilon$.

Please note that each of the following functions operates with some given function $F,$ whose evaluation could have some arbitrary time complexity; as such, our unit of time for this section is ``calls to $F$.''  Where there are terms in the complexity of a tool that do not depend on $F$'s running time, the function $t(F),$ denoting $F$'s running time, will appear in the analysis of the tool.

\subsection{Checking for a solution to $F,$ \findsol} \label{findsol}
\begin{thm}
  Take a function $F$ over a domain of size $N$.  The following algorithm {\bf findsol} determines whether there is a solution $x$ in the domain such that $F(x)=1,$ in $O(\sqrt{N/M}+\sqrt{N\lg\epsinv}M^{-1.86})$ calls to $F$ on average when there are $M$ solutions, and in $O(\sqrt{N\lg\epsinv})$ calls to $F$ on average when there are none.  If there are solutions, \findsol\ returns a random one with probability $>1-.5M^{-1.86}\epsilon$; if there is no solution or if it fails, it returns the special value {\it false} after $O(\sqrt{N\lg\epsinv})$ calls to $F$.
\end{thm}

In the following we use an extra parameter $r$, which we could never quite find a use for in the remainder of our paper.  We include it as a parameter here in case someone else is subject to greater inspiration.

The principle we use here is very straightforward.  First, we acknowledge that we can't do any better than $\sqrt{N\lg\epsinv}$ (a single \bcwz) in the case where there are no solutions, so we try to optimize for the case where there are solutions and we can hope for $O(\sqrt{N/M})$ calls to $F$.  To do this, we try \bbht\ first, due to its faster running time.  Then if we have not found a solution, we check for one with \bcwz\ to make sure.

\begin{enumerate}
\item \label{verify_bbht} Run \bbht\ up to $r$ times.  If any of those returns a result that satisfies $F,$ immediately return that result.
\item \label{verify_bcwz} Run \bcwz\ with parameter $\epsinv$.  If it returns a result that satisfies $F,$ return that result; otherwise return {\it false}.
\end{enumerate}

The analysis for this is very straightforward.
If there are solutions, step \ref{verify_bbht} takes an average of $O(2\sqrt{N/M})$ calls to $F$ (it repeats less than twice on average).  That fails with probability $O(.5^rM^{-.93r})$; if it does we move on to step \ref{verify_bcwz}, which takes $\sqrt{N\lg\epsinv}$ calls to $F$.  This gives us a total of $O(2\sqrt{N/M}+.5^rM^{-.93r}\sqrt{N\lg\epsinv})$ average calls to $F$ in the case where there are solutions; these reduce to to the promised quantities when $r=2$.
If there are no solutions, step \ref{verify_bbht} is $O(r\sqrt{N})$ and step \ref{verify_bcwz} is $O(\sqrt{N\lg\epsinv})$.

Looking at the probability of failure, we observe that the algorithm cannot possibly find a solution that does not exist, and therefore cannot fail when there are no solutions.  If there are solutions, the probability of failure is $\leq .5^rM^{-.93r}\epsilon$, the probability that the \bbht s and \bcwz\ all fail.

We chose $r=2$ because 2 is the smallest value that gives us a probability of error proportional to less than $M^{-1},$ and thus it typically minimizes running time given that condition.  Almost any constant is a reasonable choice for $r$.

\subsection{Minimum finding, \minfind} \label{minfind}

\begin{thm} \label{minfind_thm}
  Take a function $F$ over a domain of size $N$.  The following algorithm {\bf minfind} finds $x$ in the domain such that $F(x)$ is minimized, in expected time $O\left(\sqrt{N\lg\epsinv}\right)$ and with probability $\epsilon$ of failure.
\end{thm}

This algorithm is based on one by D\"urr and H\o yer\cite{dh}.  The motivation for this algorithm, as with theirs, is repeatedly to find $y$ with smaller and smaller values for $F(y)$.  To do this efficiently, we use \findsol\ as introduced in section \ref{findsol}.

\begin{enumerate}
\item Pick $y$ uniformly at random from the domain of $F$.
\item Repeat the following until instructed to return:
  \begin{enumerate}
  \item Run \findsol\ with parameter $\epsinv$ to find an element $y' : F(y')<F(y)$.
  \item If \findsol\ returns an element, set $y=y'$; otherwise return $y$.
  \end{enumerate}
\end{enumerate}

D\"urr and H\o yer show that the probability of reaching the $k\ith$ lowest value is $1/k,$ and that for different $k,$ those probabilities are independent.  With that in mind, we can sum over all values of $k$ to arrive at an average running time and a probability of failure.  For running time, we find:
\begin{align*} 
t_{\textit{minfind}}&=\sqrt{N\lg\epsinv} + \sum_{k=2}^N\frac1k \sqrt{N\lg\epsinv}k^{-1.86}\\
&\leq \sqrt{N\lg\epsinv} + \int_1^N\frac{dk}{k} \sqrt{N\lg\epsinv}k^{-1.86}\\
&\leq \sqrt{N\lg\epsinv} + \sqrt{N\lg\epsinv}
\end{align*}
calls to $F$.  We calculate the probability of failure similarly, first noting that $P_{\textit{fail}} \leq \sum_k P(k)P_{\textit{fail}}(k)$:
\begin{align*}
P_{\textit{fail}} \leq \sum_{k=2}^N\frac1k\epsilon k^{-1.86} \leq  \int_1^N\frac{dk}{k}\epsilon k^{-1.86} \leq  \epsilon
\end{align*}

\subsection{Finding all $x$ that satisfy $F,$ \findall} \label {findall}
\begin{thm}
Take a binary function $F$ over a domain of size $N,$ in which there are $M$ different parameters (solutions) that satisfy $F$.  The following algorithm {\bf findall} finds all $x$ for which $F(x)=1,$ in $O(\sqrt{NM} + \sqrt{N\lg\epsinv})$ calls to $F$ on average, with probability $\epsilon$ of failure.
\end{thm}

The idea behind this algorithm is to find successive solutions $x,$ striking each off the search as we find it in order to guarantee that we find something different every time.  We do this straightforwardly with \findsol.

\begin{enumerate}
\item Create a hash table $H$ to store results found so far.
\item Repeat the following until instructed to return:
  \begin{enumerate}
  \item \label{findall_loopstart} Run \findsol\ with parameter $\epsinv$ to find an element that satisfies $F$ but is not in $H$ (has not been found yet).
  \item If \findsol\ returns an element, add it to the result set and $H$; otherwise, return the result set.
  \end{enumerate}
\end{enumerate}

We calculate the running time with a straightforward integral.
\begin{align*}
t_{\textit{findall}}&=\sqrt{N\lg\epsinv} + \sum_{k=1}^M\left(\sqrt{N/k} + k^{-1.86}\sqrt{N\lg\epsinv})\right)\\
&\approx 2\sqrt{N\lg\epsinv} + \int_1^M dk \left(\sqrt{N/k} + k^{-1.86}\sqrt{N\lg\epsinv})\right)\\
&\approx 2\sqrt{N\lg\epsinv} + \sqrt{NM} + \sqrt{N\lg\epsinv}
\end{align*}
calls to $F$.  We calculate the probability of failure similarly, noting that $P_{\textit{fail}} \leq \sum_k P_{\textit{fail}}(k)$:
\begin{align*}
P_{\textit{fail}} \leq & \sum_{k=1}^M \epsilon k^{-1.86} \leq \int_1^M dk \epsilon k^{-1.86} \leq \epsilon
\end{align*}

Hash tables, while a useful construct, are a somewhat thorny topic in algorithms: specifically, for any hash function there is some sequence of objects to be hashed that leads to repeated collisions, causing bad asymptotic behaviour.  In cases where \findall\ will be called multiple times, as in section \ref{bfs}, in order to avoid the difficulties associated with using a hash table we can replace $H$ here with a simple array.  The initialization time for the array is $O(\widetilde{N})$ where $\widetilde{N}$ is the largest value of $N$ with which \findall\ will be called.  Every time we run \findall\ we fill $H$ up in the obvious way, keeping track of which entries we filled up in a queue and then wiping them after.

\subsection{Finding a minimal $d$ objects of different types, \mindiff} \label{mindiff}

Suppose that we want to book $d$ holidays to different destinations, and there are $N$ flights $y_i$ leaving our home airport to various destinations $G(y_i),$ with various costs $F(y_i)$.  The following algorithm finds us the $d$ cheapest destinations, and their respective cheapest flights.

\begin{thm} \label{mindiff_theorem}
  Take a function $F$ over a domain of size $N,$ and another function $G$ over the same domain.  The following algorithm {\bf mindiff} finds $d$ elements of the domain $x_i$ such that $F(x_i)$ is minimized given that all $G(x_i)$ are distinct.  More formally, given the result set of \mindiff, $x_i,$ there exists no $y$ that can ``improve'' the result set, by meeting either of the following conditions:
  \begin{enumerate}
  \item \label{mindiff_condition1} $F(y) < F(x_i)$ and $G(y) = G(x_i)$ for some $i$.  This means flight $y$ goes to $G(x_i)$ and is cheaper than $x_i$.
  \item \label{mindiff_condition2} $F(y) < F(x_i)$ for some $i,$ $G(y) \neq G(x_j)$ for any $j$.  This means $G(y)$ is a cheaper destination than one of the $G(x_i)$ --- actually it means that $y$ is a cheaper flight than the cheapest flight we've seen so far that goes to $G(x_i)$.
  \end{enumerate}
  \mindiff\ achieves this in $O\left(\left(t(F)+t(G)\right)\left(\sqrt{Nd} + \sqrt{N\lg\epsinv}\right) + d\lg N\lg d \right),$ with probability $\epsilon$ of failure.
\end{thm}

The basis for this algorithm comes from D\"urr, Heiligman, H\o yer and Mhalla\cite{dhhm}, who in their paper outline a procedure that we expound in step \ref{mindiff_them} below.  The principle behind both this algorithm and theirs is repeatedly to find $y$ such that it meets either of the conditions above, and to replace the appropriate element of the result set with the new $y$.

\begin{enumerate}
\item \label{mindiff_fictitious} Let $x$ be the array of answers.  Initially, let the $x[i]$ be ``infinities,'' for which $F(x[i])=\infty,$ and $G(x[i])$ is unique and not equal to $G(y)$ for any $y$ in the domain of $F$ and $G$.
\item Let $H$ be a hash table mapping $G(x[i])$ to $i,$ and initialize it as such.  Let $T$ be a balanced binary search tree containing the pair $(F(x[i]), i)$ for all $i,$ sorted by $F(x[i]),$ and initialize it as such.
\item \label{mindiff_them} Repeat the following until $F$ has been evaluated $O(\sqrt{Nd})$ times, or the loop has repeated $O(d \lg N)$ times (whichever happens first):
  \begin{enumerate}
  \item Let $\tau$ be the largest $F(x[k])$ in $T,$ and $k$ the corresponding index.
  \item \label{mindiff_conditionstep} Use \bbht\ to find some element of the domain $y$ such that either $F(y) < \tau$ and $G(y)\notin H$ (condition \ref{mindiff_condition2}), or $G(y)\in H$ and $F(y) < F(x[H(G(y))])$ (condition \ref{mindiff_condition1}).  Note that $F(x[H(G(y))])$ is the cost of the cheapest flight that we have found so far going to $y$'s destination, if that is currently in our result set.
  \item If condition \ref{mindiff_condition1} was met, set $x[H(G(y))] = y,$ and update $H$ and $T$ correspondingly.  Otherwise, if condition \ref{mindiff_condition2} was met, set $x[k] = y,$ and update $H$ and $T$ accordingly.
  \end{enumerate}
\item Run \findsol\ with parameter $\epsinv$ to check whether there is still a $y$ that satisfies either condition as outlined in step \ref{mindiff_conditionstep}.  If not, return $x$.  If so, repeat step \ref{mindiff_them}.
\end{enumerate}

Terminating the loop in step \ref{mindiff_them} after $O(\sqrt{dN})$ calls to $F$ provides probability of success $>\frac12,$ which is shown by D\" urr, Heiligman, H\o yer and Mhalla.  They also show that $O(d)$ iterations suffice to eliminate a constant fraction of the domain from consideration, thus $O(d\lg N)$ iterations will also provide probability of success $>\frac12$.  In order to improve the probability of success, we run \findsol\ with parameter $\epsinv$ to check whether we are yet done; if we are not, we repeat step \ref{mindiff_them} until we are.  Since the probability for step \ref{mindiff_them} to finish successfully after one pass is $\geq\frac12,$ we expect to repeat it -- and \findsol\ -- an average of $\leq 2$ times.  We also have to consider the contribution of updating and accessing $T,$ which will take $O(\lg d)$ time with every iteration; thus our total running time is $O\left(\left(t(F)+t(G)\right)\left(\sqrt{dN} + \sqrt{N\lg\epsinv}\right)+ d\lg N\lg d\right)$ with probability $1-\epsilon$ of success.

Note that if $d$ is greater than the number of distinct values for $G$ ($\equiv \gamma$), we return $\gamma$ valid elements and $d-\gamma$ infinities (fictitious elements of the domain as defined in step \ref{mindiff_fictitious}).

As with \findall, we use a hash table here that can be replaced by an array if \mindiff\ is going to be used multiple times.

\section{Graph algorithms} \label{graph}

A graph is a mathematical construct made up of a set of vertices $v_a,$ and a set of edges $e_{ab}$ that connect the vertices together.  Typically one thinks of the vertices as locations and the edges as connections between them: for example, one could represent bus stops in a city as the vertices of a graph, and the paths of buses as the edges connecting them.  Graphs are widely applicable throughout the field of algorithms, sometimes showing up in unexpected places as useful constructs to solve problems.

Each edge in a graph connects two vertices $v_a$ and $v_b,$ and is either {\it directed} ($v_a\rightarrow v_b$) or {\it undirected} ($v_a\leftrightarrow v_b$); typically graphs contain only directed or only undirected edges.  In a {\it weighted} graph edges have some weight associated with them, typically thought of as a cost or distance associated with moving from $v_a$ to $v_b$ (and vice-versa in the undirected case).  An unweighted graph can be thought of as a weighted graph whose edge-weights are all 1.

With the concept of edges having some cost or length, we can discuss problems such as shortest paths: given a graph, what is the ``shortest'' path -- the path of minimal summed length -- from some source vertex to some destination vertex, or possibly to every destination vertex?  Suppose we want the shortest paths from every vertex to every other vertex: can we calculate them faster than we can by running our single-source shortest paths algorithm from each source?  What if some of the edges have negative weights: are our algorithms affected?

In this section we will focus on quantum versions of long-studied classic problems such as shortest paths, searching through graphs, and graph matchings (suppose you want to pair up vertices that are connected; what's the maximum number of pairs you can make?).

We present the algorithms here for two models of representing graphs, both of which we will assume are given to us as quantum black boxes.  In both models, $V$ is the number of vertices and $E$ the total number of edges in the graph; $\mathbb{V}$ and $\mathbb{E}$ represent the vertex set and edge set respectively.  If there is an edge between vertices $v_i$ and $v_j,$ we refer to it as $e_{ij}$.  The models are:
\begin{itemize}
\item The {\bf adjacency matrix} model, as a quantum black box, is passed $i,j$ ($0\leq i,j < V$) and returns whether $e_{ij}$ exists.  Conceptually this could be determined by some mathematical function, but classically the graph is usually represented as a $V\times V$ matrix with entries in $\{0,1\}$.
\item The {\bf edge list} model, as a quantum black box, is passed $i,j$ and returns the destination of the $j^{\text{th}}$ edge outgoing from vertex $v_i$ (we assume for convenience that we know how many edges are outgoing from each vertex).  Classically this is usually represented as a ragged array, but sometimes is generated mathematically as-needed.  We call the set of edges outgoing from $v_i$ $d[i],$ and its cardinality $\modd{i}$.  The edge list model is sometimes called the {\it adjacency array} model.
\end{itemize}

If the graph is weighted, the adjacency matrix and edge list models also return the weight of the edge queried.

For an excellent resource on graph theory and algorithms therein, please see Cormen, Leiserson, Rivest and Stein's classic introduction to algorithms\cite{clrs}.  It contains detailed discussions of breadth-first and depth-first searches, Dijkstra's algorithm and the Bellman-Ford algorithm, as well as all-pairs shortest paths.  We look at all of these in this section, but leave the details to this reference.

In this section, we assume that the desired probability of failure $\epsilon$ is such that $\epsinv$ is polynomial in the number of vertices $V$.  Note that the number of edges $E$ can be no more than $O(V^2)$ for the graphs we will be discussing here (see Appendix \ref{comments}), so ``polynomial in $V$'' $\Rightarrow$ ``polynomial in $E$.''  The error analysis for this section can be found in Appendix \ref{error_graph}.

\subsection{Breadth-first search, \bfs} \label{bfs}

Breadth-first and depth-first search are two of the simplest algorithms for searching a graph, and find extensive use inside many important graph algorithms.  The principle behind each is the same: starting at some source, we systematically explore the vertices of our graph, ``visiting'' each vertex connected to the origin in some order.  By introducing quantum versions of each here, we tarnish their simplicity but maintain their strength and increase their speed.

As we mentioned above, \bfs\ and \dfs\ both see extensive use.  Both can be used to determine whether a vertex is connected to the rest of the graph, and breadth-first search in particular can be used to compute shortest paths in an unweighted graph.  Depth-first search, on the other hand, can be used to detect ``bridges'' in a graph: edges which, if they were removed, would sever the graph into two pieces with no edges between them.  There is a great deal of utility to be had from these two over and above what is discussed here, and both are very simple, solved problems in classical computing.

To implement a breadth-first search here, we take an approach based heavily on classical BFS: we keep a list of vertices we want to visit, and every time we visit another of those vertices we add all of its unvisited neighbours to the list.  Through use of a boolean array we ensure each vertex is only visited and added once.  To choose the order in which the vertices are visited, we let our list be a ``queue,'' wherein vertices added first are visited first; thus we end up visiting the vertices in order of how close they are to the origin of our search (breadth-first).  To speed up the process of finding all of the unvisited neighbours of each node, we use section \ref{findall}'s \findall.  This algorithm is based on a BFS from Ambainis and \v{S}palek\cite{as}, though they use repeated \bbht s rather than our \findall.

\begin{thm}
The following algorithm {\bf BFS} executes a breadth-first search through a graph \ggdef\ in $O(\sqrt{V^3\lg V})$ time in the matrix model, $O(\sqrt{VE\lg V})$ in the edge list model.
\end{thm}
\begin{enumerate}
\item Let the vertex from which we are searching be called $v_a$.  Let there be a queue of vertices $q,$ and let it initally contain only $v_a$.  Let there be a boolean array $vis$ of size $V,$ with entries $vis[i]=\delta_{i,a}$.
\item Repeat the following until $q$ is empty:
  \begin{enumerate}
  \item Remove the first element of $q$ and call it $v_i$.
  \item Visit $v_i$.
  \item \label{bfs_findall} Using section \ref{findall}'s \findall, find all neighbours $v_j$ of $v_i$ with $vis[j]=\textit{false}$.
  \item For each such $v_j,$ set $vis[j]=true$ and add $v_j$ to $q$.
  \end{enumerate}
\end{enumerate}

In the matrix model, each vertex $v_i$ is processed at most once and contributes $\sqrt{Vn_i} + \sqrt{V\lg V},$ where $n_i$ is the number of elements added to $q$.  In the edge list model, each vertex is processed at most once and contributes $\sqrt{\modd{i}n_i} + \sqrt{\modd{i}\lg\modd{i}}$.  By the Cauchy-Schwartz inequality, we have:
\begin{equation}
  \sum_{v_i\in\mathbb{V}}{\sqrt{n_i\modd{i}}} \leq \sqrt{\sum_{v_i\in\mathbb{V}}{n_i}}\sqrt{\sum_{v_i\in\mathbb{V}}{\modd{i}}} \leq \sqrt{VE}
\end{equation}
\begin{equation} \label{bfssum}
  \sum_{v_i\in\mathbb{V}}{\sqrt{\modd{i}\lg\modd{i}}} \leq \sqrt{\sum_{v_i\in\mathbb{V}}{\modd{i}}}\sqrt{\sum_{v_i\in\mathbb{V}}\lg\modd{i}} \leq \sqrt{VE\lg V}
\end{equation}
Thus \bfs\ in the edge list model runs in $O(\sqrt{VE\lg V}),$ and since $E<V^2,$ \bfs\ in the matrix model runs in $O(\sqrt{V^3\lg V})$.  Classically breadth-first search takes $O(E)$ time, so \bfs\ is faster than its classical counterpart for $E\in\Omega(V\lg V)$.

\subsection{Depth-first search, \dfs} \label{dfs}

Classically, depth-first and breadth-first search can have very similar implementations, and the same is true in the quantum regime.  The simplest implementation of depth-first search in both regimes, however, is a recursive one, which we show here.

\begin{thm}
The following algorithm {\bf DFS} executes a depth-first search through a graph \ggdef\ in $O(\sqrt{V^3\lg V})$ time in the matrix model, $O(\sqrt{VE\lg V})$ in the edge list model.
\end{thm}

\begin{enumerate}
\item Let the vertex from which we are searching be called $v_a$.  Let there be a boolean array $vis$ of size $V,$ with entries $vis[i]=0$.  Call $\text{DFS-BODY}(v_a)$.
\item Function {\bf DFS-BODY(vertex $v_k$):}
  \begin{enumerate}
  \item Visit $v_k$.  Set $vis[k] = \textit{true}$.
  \item \label{dfs_findsol} Use section \ref{findsol}'s \findsol\ to find a neighbour of $v_k$ that has not yet been visited, $v_i$.
  \item If there is some such $v_i$:
    \begin{enumerate}
    \item Recursively call $\text{DFS-BODY}(v_i)$.
    \item After returning from the recursive call, go back to step \ref{dfs_findsol}.
    \end{enumerate}
  \end{enumerate}
\item Return.
\end{enumerate}

There are two contributions to our running time here, which we will work through in the edge list model.  The first is that for each vertex visited, \findsol\ must fail once, leaving us with a contribution of $O(\sqrt{VE\lg V})$ (see equation \ref{bfssum}).  The second contribution is the sum of the running times of the successful \findsol s.  We sum again over vertices, noting that for each vertex $v_i,$ if we end up finding $n_i$ of its neighbours through $\text{DFS-BODY}(v_i),$ the running time of that will be $O\left(\sum_{k=1}^{n_i} \sqrt{(\modd{i}/k)\lg\modd{i}}\right),$ and therefore $O(\sqrt{\modd{i}n_i\lg\modd{i}})$.  Summing that contribution over each vertex, we again arrive at $O(\sqrt{VE\lg V})$ through equation \ref{bfssum}.  In the matrix model we simply replace $E$ with $V^2,$ arriving at $O(\sqrt{V^3\lg V})$.

Classically depth-first search takes $O(E)$ time, so \dfs\ is faster than its classical counterpart for $E\in\Omega(V\lg V)$.

\subsection{Single-source shortest paths with negative edge weights, \spnw} \label{spnw}

The problem of single-source shortest paths, finding the shortest paths through a graph from some source $v_a$ to all destinations, is solved elegantly by D\"urr, Heiligman, H\o yer and Mhalla\cite{dhhm} with an algorithm loosely based on Dijkstra's; their algorithm does not allow negative edge weights, so here we base an algorithm on Bellman-Ford, which does\cite{kholondyrev,bellman,ford}.  Our algorithm returns an array of shortest distances to points, or the special value {\it false} if there exists a negative-weight cycle in the graph that can be reached from the source.  It also computes an array {\it from}, whose $i\ith$ element is the index of the vertex previous to $v_i$ on the shortest path from $v_a$ to $v_i$; this allows the shortest path from $v_a$ to $v_i$ to be recovered.

Intuitively, we are going to take each edge in turn and see if it helps our current shortest path to each point; we repeat that process $V$ times, at which point each edge will have helped all it can.

\begin{thm}
Given a graph \ggdef, the following algorithm {\bf SPNW} returns an array whose $i\ith$ element is the shortest distance from the source $v_a$ to vertex $v_i,$ $\infty$ if no such path exists.  If there is a negative weight cycle that can be reached from $v_a,$ instead of an array it returns the special value {\it false}.  It does this in $O(\sqrt{V^5\lg V})$ time in the matrix model, $O(\sqrt{V^3E\lg V})$ in the edge list model.
\end{thm}

\begin{enumerate}
\item If we are using the edge list model, set up an array $f$ such that $f[i][j]$ is the source of the $j\ith$ edge incident on $i$.
\item Initialize an array $dist,$ such that $dist[i]=\infty$ for $i\neq a,$ $0$ for $i=a$.
\item Initialize an array {\it from}, such that $\textit{from}[i]=-1$.
\item \label{spnw_loop} Repeat the following $V-1$ times:
  \begin{enumerate}
  \item \label{spnw_assign} For each vertex $v_i,$ using the algorithm of section \ref{minfind}, \minfind\ a vertex $v_j$ such that $e_{ji}$ exists, and $dist[j] + \length(e_{ji})$ is minimized.  Execute the minfind by searching over $f[i]$ in the edge list model, $\mathbb{V}$ in the matrix model.
  \item If $dist[j] + \length(e_{ji}) < dist[i],$ set $dist[i]=dist[j] + \length(e_{ji})$ and set $\textit{from}[i] = j$.
  \end{enumerate}
\item Repeat step \ref{spnw_assign} one more time.  If it changes $dist,$ return {\it false}.  Otherwise return $dist$.
\end{enumerate}

This algorithm, like Bellman-Ford, works due to the fact that all shortest paths in a graph without negative weight cycles must use fewer than $V$ edges.  Each time through step \ref{spnw_loop}, we ask ``could the path to vertex $v_i$ be shorter if we were allowed to use one more edge?''  Repeating this $V-1$ times lets us use $V-1$ edges, and repeating it a last time lets us check whether there is a negative weight cycle.  Meanwhile we keep our array {\it from}, which tells us how we got to $v_i$ and allows us to recover the whole path.  In the edge list model, the running time is $V\sum_i{\sqrt{\modd{i}\lg\modd{i}}} = O(\sqrt{V^3E\lg V})$ by equation \ref{bfssum}.  In the matrix model, our $E$ becomes a $V^2$ as usual, and we have $O(\sqrt{V^5\lg V})$.  Note that since this is greater than $V^2,$ if the graph is sparse it may be worth first converting to the edge list model.

Classically single-source shortest paths with negative edge weights takes $O(VE)$ time, so \spnw\ is faster than its classical counterpart for $E\in\Omega(V\lg V)$.

\subsection{All-pairs shortest paths with negative edge weights, \apsp} \label{apsp}

\begin{thm}
Given a graph \ggdef, the following algorithm {\bf APSP} returns an array whose $i,j^{\text{th}}$ element is the length of the shortest path between vertices $v_i$ and $v_j,$ $\infty$ if no such path exists.  If there is a negative weight cycle in the graph, instead of an array it returns the special value {\it false}.  It does this in $O(\sqrt{V^5}\lg V)$ in the matrix model, $O(\sqrt{V^3E}\lg V + V^2\lg^3 V)$ in the edge list model.
\end{thm}

We can do this directly with Johnson's algorithm\cite{clrs,johnson,dustinmikematthew}.  Johnson's works by running Dijkstra's algorithm from every origin point, which gives the shortest paths from all points to all other points; the difficulty is that Dijkstra's does not work in graphs with negative-weight edges, so first it is necessary to reweight edges so that all of their weights are positive.  That is accomplished through the application of a single Bellman-Ford, which also tells us whether there are any negative-weight cycles in the graph.

In our quantum version, we alter Johnson's by replacing its call to Bellman-Ford with a call to section \ref{spnw}'s \spnw, and its calls to Dijkstra's algorithm with calls to our modification of D\"urr, Heiligman, H\o yer and Mhalla's single-source shortest paths (section \ref{sssp}, \cite{dhhm}).  The \spnw\ serves to reweight the edges so that they are all positive, and then we run single-source shortest paths from each vertex.  Our total complexity is the sum of $V$ single-source shortest paths and one \apsp, which totals to $O(\sqrt{V^5}\lg V)$ in the matrix model, $O(\sqrt{V^3E}\lg V + V^2\lg^3 V)$ in the edge list model.

Classically all-pairs shortest paths with negative edge weights takes $O(VE + V^2\lg V),$ so \apsp\ is better than its classical counterpart for $E\in \Omega(V\lg^3V)$.  There is another classical algorithm, by Zwick\cite{zwick}, which runs in $O(V^{2.575})$; \apsp\ is asymptotically better than Zwick's algorithm in the worst case.

\section{Improvements to existing quantum graph algorithms} \label{other_papers}

It has quickly become to the tradition in the literature\cite{dhhm,as} to devise quantum algorithms with \bbht\ as though there were no probability that it could fail, and then to throw a factor of $\log(N)$ into the running time at the end to take the probability of failure into account.  Here we give two examples of algorithms that can be given faster asymptotic behaviour with careful error analysis.

\subsection{Single-source shortest paths} \label{sssp}

D\"urr, Heiligman, H\o yer and Mhalla\cite{dhhm} discuss algorithms for single-source shortest paths, minimum spanning tree, connectivity and strong connectivity.  The quantum query complexity for their single-source shortest paths, $O(\sqrt{VE}\lg^2V),$ can be improved by using \mindiff, whereupon it becomes $O(\sqrt{VE}\lg V)$.  The explanation follows, and is best enjoyed with their paper in hand.

Step 2(a) in their algorithm involves using what we have called \mindiff\ (see section \ref{mindiff}).  Their version of it runs in $O(\sqrt{Nd})$ queries to the graph and with constant probability of failure; they repeat this $\log N$ times on every call to reduce the probability of failure to $1/N$.  We use our \mindiff\ with $F(e_{ij})=\text{length}(e_{ij}),$ $G(e_{ij})=j$ instead, which runs in $\sqrt{Nd}+\sqrt{N\lg\epsinv}$ queries to the graph.

Summing as they do to compute running time (in their notation where $n=V, m=E$) we have: $\sum_{j=1}^{n/s}\left(\sqrt{sm_j} + \sqrt{m_j\lg\epsinv} + s\lg m_j\lg s \right),$ which by the Cauchy-Schwartz inequality (and some algebra on the last term) is $\leq \sqrt{(s)(n/s)(m)} + \sqrt{(n/s)(m)(\lg\epsinv)} + n\lg s\lg(ms/n)$ which is of order $\sqrt{nm}\left(1+\frac{\sle}{s}\right) + n\lg s\lg n$.  Summing over sizes, where $s=1,2,4,\ldots n,$ we arrive at $\leq \sqrt{nm}(2\lg n + 2\sle)+n\lg^3 n,$ which is (returning to our notation) $O(\sqrt{VE}\lg V + V\lg^3 V)$.

D\"urr, Heiligman, H\o yer and Mhalla do not make some specifics of their version of \mindiff\ clear, such as how they maintain the list of their best answers so far.  This will inevitably add to the total running time of their algorithm (though not its queries to the graph, which is what they chose to analyze), and so their total running time ends up as $O(\sqrt{VE}\lg^2 V + ?)$.

Our total complexity has to include their step 2(b), finding the minimum element of all the $A_i$ whose $v$ is not in any $P_i$.  This can be done by keeping a balanced binary search tree $T$ with average $O(\lg N )$ insertion/removal/access, which contains all such $A_i$.  Every time a $P_i$ of size $s$ is changed, we remove the old elements from $T$ and insert the new ones.  This runs in $s\lg V$ every time we change a $P_i$ of size $s,$ and each size is created/destroyed no more than $V/s$ times, for a total of $V\lg V$ for each size.  Summing over the $\lg V$ different sizes, we arrive at $V\lg^2V$.  Thus our total complexity remains $O(\sqrt{VE}\lg V + V\lg^3V)$.

The best classical solution to this problem, Dijkstra's algorithm, runs in $O(E + V\lg V),$ so the quantum algorithm is better for $E\in\Omega(V\lg^3 V)$.

\subsection{Bipartite matching} \label{bipartite}

Ambainis and \v{S}palek\cite{as} address bipartite matching, non-bipartite matching and maximum flow.  Their algorithm for bipartite matching takes $O(V\sqrt{E+V}\lg V)$ time, and is a quantum adaptation of Hopcroft and Karp's classical $O((E+V)\sqrt{V})$ algorithm\cite{hk}; we solve the problem here in $O(V\sqrt{(E+V)\lg V})$.

The problem of bipartite matching can be described in several ways: for example, consider a collection of boys and girls to be vertices of a graph, and have an edge in the graph for each $(boy, girl)$ pair that would make a good couple.  In bipartite matching, we pair off the boys and girls in such a way that only compatible couples are paired, each person has at most one partner, and there is a maximum number of pairings.

Some basic principles underlie most solutions to this problem.  Consider some (non-maximum) matching-so-far $\mathbb{M}$ between boys and girls; if we can construct a path $\mathbb{P}$ starting at an unmatched boy and ending at an unmatched girl such that all edges in the path are either unused $boy\rightarrow girl$ edges or used $girl\rightarrow boy$ edges, then the old matching can be expanded by 1 more pair by taking $\mathbb{M}' = \mathbb{M}\oplus\mathbb{P}$ (where $\mathbb{M}\oplus\mathbb{P}$ means taking all edges in either $\mathbb{M}$ or $\mathbb{P},$ but not both).  Intuitively, where $\mathbb{M}$ and $\mathbb{P}$ have an edge in common, we are ``unmatching'' that $(boy, girl)$ pair, and ``rematching'' the two using the surrounding edges in the path.  Because this path augments $\mathbb{M}$ by adding one to its size, it is called an ``augmenting'' path.

The principle behind Hopcroft and Karp's algorithm is as follows: suppose that every time we want to find an augmenting path $\mathbb{P},$ we find the shortest such path.  They proved that if we do that, we will see at most $2\sqrt{V}$ different path lengths in the whole process of constructing a maximum matching.  So if we devise a process to find a {\it maximal} set of augmenting paths of minimal length, (maximal means here that the set cannot be expanded by adding more paths of the same length) we can repeat that process $O(\sqrt{V})$ times and have constructed a maximum matching.

The construction of a maximal set of augmenting paths of minimal length is accomplished through the use of a breadth-first search and a depth-first search, the details of which we leave to our references.  They can however be replaced by our \bfs\ and \dfs\ functions, giving us a total running time of $O(V\sqrt{E\lg V}),$ a whopping $\sqrt{\lg V}$ faster than Ambainis and \v{S}palek's algorithm.  This is also faster than the classical solution, when $E\in\Omega(V\lg V)$.

Ambainis and \v{S}palek also discuss non-bipartite matching and maximum flow in the same paper; in both cases they ignore errors for the body of their algorithms, and throw on an extra factor of $\log V$ at the end in order to reduce the probability of failure to a constant.  While that works, this section shows that it is not necessarily optimal for bipartite matching; and due to the similarity of bipartite matching to the other problems they consider, it is reasonable to guess that one could also achieve an $O(\sqrt{\log V})$ speedup for general matching and flow.

\section{Computational geometry algorithms}

Geometry problems are a natural area of attack for quantum algorithms, because by defining $N$ points we have implicitly defined $O(N^2)$ relationships between those points, making it very natural to ask questions whose answers require information $O(N^2)$ in the size of the question.  We will address points as $p_i$.

In this section, we make reference to the probability of error $\epsilon$ but do not discuss it in depth.  The error analysis for this section can be found in Appendix \ref{error_geometry}.

\subsection{Maximum points on a line, \maxpoints} \label{maxpoints}

This problem is, in all of its generality, a very simple one: given $N$ points, find the line that goes through the maximum number of them.  We differentiate here between a solution that is practical for integers\cite{stevens} and a slightly slower solution that is practical for real numbers; acknowledging that practical computers, however quantum, do not offer consistent, identical normalization for parallel vectors of real numbers.

Intuitively each algorithm works by taking a single point $p$ and finding out how many points are on the best line that goes through $p$.  We then use \minfind\ to find the best such $p$.  In the $\mathbb{Z}^n$ case, our method is to find the vector from $p$ to each other point, canonicalize it using GCD, and then stick all those vectors into a hash table so that we can quickly count repeats.  In the $\mathbb{R}^2$ case, our method is to sort the points in counterclockwise order about $p$ and see look for collinear points, which should now be ordered consecutively.

This is a particularly interesting problem to solve in $\mathbb{Z}^2$ because it is a member of a class of classical problems called ``3SUM-hard''\cite{go}.  Of the problems belonging to this class, all of the known ones have classical lower-bounds of at most $\Omega(N),$ and upper bounds of at least $O(N^2)$.  All problems in the class reduce to the 3SUM problem: given a set $S$ of $N$ integers, is there some triplet $a,b,c$ in that set such that $a+b+c=0$?  This is quite a straightforward problem to solve with \findsol\ in $O(N),$ while we will solve this problem in $N^{1.5},$ opening a gap of $N^{.5}$ between two similar problems, where no such gap existed before.  This raises interesting questions about the maximum points on a line problem, and a number of other problems in 3SUM-HARD.
which in turn suggests that many of the algorithms in 3SUM-hard (such as \maxpoints) may be amenable to $\text{sub-}N^2$ solutions.

\subsection{Maximum points on a line: $\mathbb{Z}^n$} \label{maxpoints_zn}

\begin{thm}
Let there be $N$ points in $\mathbb{Z}^n,$ whose coordinates are bounded by $\pm U$.  The following algorithm {\bf maxpoints} finds the straight line on which lies the maximum number of those points, in $O(N^{3/2}n\lg U\sle)$ time and with probability of $\epsilon$ of failure.
\end{thm}

\begin{enumerate}
\item Use section \ref{minfind}'s \minfind\ to maximize the following function, $mup$ (maximum using $p$), over all points $p$.  Call the result $P$.
\item Function {\bf mup}:
  \begin{enumerate}
  \item Create an empty hash table $H,$ mapping vectors in $\mathbb{Z}^n$ (keys) to integers (values).
  \item For each point $p_i$:
    \begin{enumerate}
    \item Define $\vect{a} = \vect{p_i} - \vect{p}$.
    \item Normalize $\vect{a},$ keeping its entries in the integers, so that the first nonzero component is positive and the gcd of the absolute values of the components is 1.
    \item If $\vect{a}$ is not yet in $H,$ insert it in $H$ mapping to value 1; if $\vect{a}$ is already in $H,$ increment its value.
    \end{enumerate}
  \item Return the maximum value in $H$: the number of points on the best line going through $p$.
  \end{enumerate}
\item Run $mup$ on $P,$ but instead of returning the maximum value in the hash table return its corresponding key, and call it $\vect{V}$. 
\item The answer to return is the line $\vect{X(t)} = \vect{P} + t\vect{V}$.
\end{enumerate}

In $mup,$ all vectors to other points from $p$ are canonicalized in such a way that any pair of points collinear with $p$ will have the same direction vector $\vect{a}$.  $mup$ repeats $n$ {\it gcd}s $N$ times, for a total of $O(Nn\lg U),$ and our main function's most costly operation is one \minfind\ that evaluates $mup$ $O(\sqrt{N\lg\epsinv})$ times.  Thus our total running time is $O(N^{3/2}n\lg U\sle),$ and our probability of failure is $\epsilon$.  Classically the problem can be solved in $N^2n\lg U$.

\subsection{Maximum points on a line: $\mathbb{R}^2$} \label{maxpoints_r2}

\begin{thm}
Let there be $N$ points in $\mathbb{R}^2$.  The following algorithm finds the straight line on which lies the maximum number of those points in $O(N^{3/2}\lg N\sle),$ with probability of failure $\epsilon$.
\end{thm}

\begin{enumerate}
\item Use \minfind\ to maximize the following function, {\it mup2}, over all points $p$.  Call the result $P$.
\item Function {\bf mup2}:
  \begin{enumerate}
  \item Let $\vect{a_i} = \vect{p_i} - \vect{p}$.  If $\vect{a_i}.x < 0,$ or $\vect{a_i}.x = 0$ and $\vect{a_i}.y < 0,$ then reverse $\vect{a_i}$.  This puts all points to the right of $p$.
  \item Sort the $\vect{a_i}$ as follows: $\vect{a_i} < \vect{a_j}$ iff $\left(\vect{a_i} \times \vect{a_j}\right) \cdot \widehat{z} > 0$.  This has the effect of sorting the $p_i$ in counter-clockwise order about $p$.
  \item Iterate over the sorted array, keeping a running total of how many consecutive $\vect{a_i}$ have cross product of 0 with one another.  Return the maximum such total.  (Practically, we should see how many consecutive $\vect{a_i}$ have cross product $<\delta$ for some small $\delta,$ and loop through a second time to catch the nearly-straight-up and nearly-straight-down $\vect{a_i}$).
  \end{enumerate}
\item Run $mup2$ on $P,$ but instead of returning the maximum total, return some point (other than $P$) on the line giving that total.  Call it $P'$.
\item The answer to return is the line $\vect{X(t)} = \vect{P} + t(\vect{P'} - \vect{P})$.
\end{enumerate}

This algorithm sorts the points about each point $p,$ which has the effect of grouping collinear points together.  Then it simply counts how many consecutive collinear points it can find.  {\it mup2} is $O(N\lg N),$ and our most costly operation is one \minfind\ that evaluates $mup2$ $O(\sqrt{N\lg\epsinv})$ times, for a total running time of $O(N^{3/2}\lg N \sle)$ and probability of failure $\epsilon$.  Classically this problem can be solved in $O(N^2\lg N)$.

\section{Dynamic Programming algorithms} \label{dp}

Dynamic programming (DP) is a method that solves problems by combining the solutions to subproblems.  DP algorithms achieve this by partitioning their problems into subproblems, solving the subproblems recursively, and then combining the solutions to solve the original problem.  What distinguishes dynamic programming from other approaches is that the subproblems are not independent: subproblems share sub-subproblems with one another.  A dynamic programming algorithm solves every sub-subproblem only once and saves its result in a table, thus eliminating the need to recompute the answer for a sub-subproblem every time it is needed.

Dynamic programming is often used to solve optimization problems.  Given some situation (a problem), come up with a choice (each possible choice leads to a subproblem) that optimizes some final quantity (way down at the $\text{sub}^n\text{-problem}$ level).  We will see an example of this in section \ref{coinchange}.  Since DP is often used to make some sort of optimal choice, DP algorithms in general are obvious candidates for section \ref{minfind}'s \minfind, which square-roots the process of checking all our options.



In this section, we assume that the desired probability of failure $\epsilon$ is such that $\epsinv$ is polynomial in the size of the input.  In some places this affects the running time, and so we make reference to $\epsilon$ but do not discuss it in depth.  The error analysis for this section can be found in Appendix \ref{error_dp}.

\subsection{Coin changer, \coinchange} \label{coinchange}

Given a monetary system with some set of coins and bills, we may wish to make some precise amount of money -- the {\bf coin changer} problem is to use as few coins and bills as possible.  Intuitively, this is easy: with Canadian or American money, for example, to make $D$ cents one can simply take the largest bill/coin of value $v \leq D,$ then make $D-v$ cents in the same way.  For example, to make 40\cents\ one would take the largest coin less than 40\cents\ (25\cents), then the largest coin less than the remaining 15\cents\ (10\cents), and finally a 5\cents\ coin.  This is a {\it greedy} approach that works for most real currencies, but it is not always optimal: for example, should a 20\cents\ piece be added to the Canadian system, then making 40\cents\ only takes two coins, but the greedy approach will still cause us to use three.  Should the reader ever travel to Costa Rica or Bhutan, he or she will encounter a non-greedy currency system.

\begin{thm}
Given a length $C$ integer array of coin denominations $V,$ as well as an integer $D,$ the following algorithm {\bf coinchange} returns the minimum number of coins required to make $D$ units, or $\infty$ if making $D$ units of currency is impossible.  It achieves this in $O(D\sqrt{C\lg D})$ time.
\end{thm}

Since we are trying to minimize a quantity, the number of coins used, making $D$ units optimally is a matter of choosing one coin $V[i]$ to use, then making $D-V[i]$ units optimally.  To do so we build up a table $T,$ where $T[i]$ is the minimum number of coins needed to make $i$ units.  We start by filling in $T[i]$ with $i$ small, since later entries will depend on earlier ones.

\begin{enumerate}
\item Let there be an array $T$ of size $D+1,$ such that initially $T[0]=0,$ and $T[i\neq 0] = \infty$.
\item For $d$ from 1 to $D,$ DO:
  \begin{enumerate}
  \item \label{coinchange_minfind} Use the algorithm of section \ref{minfind} to \minfind\ one of the coins $V[i]$ such that $d - V[i] \geq 0,$ and $1 + T[d - V[i]]$ is minimal.
  \item If such a coin was found, let $T[d] = 1 + T[d - V[i]]$.
  \end{enumerate}
DONE.
\item Return $T[D]$.
\end{enumerate}

Here we simply fill in the table as discussed above, by using \minfind\ to determine which coin should be taken first.  The minfind takes $O(\sqrt{C\lg D})$ time, and is repeated $D$ times for a total time complexity of $O(D\sqrt{C\lg D})$.

The reason we discuss this example is because it is very representative of how one can improve dynamic programming algorithms in general using quantum techniques, and as such is a good forum for the discussion of quantum DP in general.  For example, many dynamic programming algorithms, including this one, have alternate recursive implementations: rather than consulting entries of a table that have already been filled in, we call our function recursively on their indices.  Rather than consulting $T[x],$ we call $mincoins(x),$ and it calls $mincoins(x-25)$ and $mincoins(x-10),$ etc.  To save ourselves from exponential repetition, whenever we compute the result for a subproblem we cache it; so that the next time $mincoins$ is called with the same parameters, we simply return the result.  The advantage of recursive DP (often called {\it memoization}) is that for many people it is very intuitive to write a recursive function that computes the result, then throw in a few lines that cache and retrieve the cached value.

Classically, memoization is valuable primarily as an alternate way of implementing dynamic programming; it is only faster in rare cases.  Indeed, many DP algorithms are more efficient (use less memory) when implemented iteratively, and some few have no clear implementation through memoization.

To implement memoization in the quantum case, one could use \findsol\ to find the subproblems whose solutions have not been cached yet, call those recursively, and then take the appropriate action, such as a \minfind\ over the subproblems.  There is no clear alternative to this approach, which is unfortunate: it can lead to asymptotically longer running times than standard DP.  This is a little tricky to prove, and somewhat outside the scope of this paper; for those who are interested, we suggest considering a carefully chosen dependency graph such that there is a set $X$ of many states with no dependencies, and an asymptotically smaller set $Y$ of states that depend on subsets of $X$ ($X$ might have size $N^6,$ $Y$ have size $N^4,$ and each element of $Y$ could depend on $N^4$ elements of $X$).

\subsection{Maximum subarray sum, \subarraysum} \label{subarray-sum}

\begin{thm}
Given an $N \times N$ array of real numbers $A,$ the following algorithm {\bf subarray-sum} finds a rectangular subarray such that the sum of the subarray's elements is maximized, in $O(N^2\sle)$ time and with probability of failure $\epsilon$.  We will address the result by its limits: $(miny,minx,maxy,maxx)$.
\end{thm}

This is another classic problem, for which the best known classical solution runs in $O\left(N^3\sqrt{\frac{\log\log N}{\log N}}\right)$ and was found by Tamaki\cite{tamaki}.  There is a much more straightforward (though still clever) $O(N^3)$ solution, which involves maximizing the sum of all $O(N^2)$ possible column ranges, each in $O(N)$.

Our algorithm begins by creating a table $T$ that makes checking the sum for an arbitrary rectangle $O(1),$ and then simply \minfind s over all rectangles.  This algorithm, like the classical one, is really greedy rather than dynamic programming; we include it in this section because the construction of $T$ is DP.

\begin{enumerate}
\item Let there be an $N\times N$ array $T,$ whose $i,j$ element will hold the sum for subarray $(0,0,i,j)$.  Initialize its entries to $0,$ and define $T[i][j]=0$ if $i$ or $j$ is negative. The next step will fill in $T$ as desired.
\item For $i$ from $0$ to $n - 1,$ For $j$ from $0$ to $n - 1$ DO:
  \begin{enumerate}
  \item $T[i][j] = A[i][j] + (T[i-1][j] + T[i][j-1] - T[i-1][j-1])$.
  \end{enumerate}
  DONE.
\item \label{subarray-sum_minfind} There are $N^4$ possible rectangular subarrays.  The summation over any such array is $T[maxy][maxx] - T[maxy][minx - 1] - T[miny - 1][maxx] + T[miny - 1][minx - 1],$ which is an $O(1)$ calculation.  Use the algorithm of section \ref{minfind} to \minfind\ over all such $(miny,minx,maxy,maxx)$ and find the subarray with the maximum summation, and then return it.
\end{enumerate}

The creation of $T$ takes $O(N^2),$ and the \minfind\ takes $O(N^2\sle)$ and has probability of failure $\epsilon$.  The dynamic programming part of this algorithm is the construction of $T,$ which could also be implemented using memoization as discussed above.

\section{Conclusions} \label{conclusions}

We summarize our results from sections \ref{tools}-\ref{dp} here.  Results from tables \ref{graph_table} to \ref{dp_table} should be checked against Appendix \ref{error} for their exact error-dependence: in the interest of brevity, we often assume the probability of error $\epsilon$ to be such that $\epsinv$ is polynomial in $N$ (or $V$), or is constant.

\begin{table}[ht!]
\begin{center}
\begin{tabular}{|lll|}
\hline
problem
& quantum complexity
& classical (avg)\\ \hline
finding one solution
& $O(\sqrt{N/M} + \sqrt{N\lg\epsinv}/M^{1.86})$
& $O(N/M)$ \\
\ \ same algorithm, no solutions
& $O(\sqrt{N\lg\epsinv})$
& $O(N)$ \\
minimum finding
& $O(\sqrt{N\lg\epsinv})$
& $O(N)$ \\
finding all $M$ solutions
& $O(\sqrt{NM} + \sqrt{N\lg\epsinv})$
& $O(N)$ \\
finding $d$ min.\ diff.\ objects
& $O(\sqrt{Nd}+\sqrt{N\lg\epsinv}+d\lg N\lg d)$
& $O(N)$ \\
\hline
\end{tabular}
\end{center}
\vskip-14pt
\caption{Tools.  The unit of time is calls to $F$}
\label{tools_table}
\vskip10pt
\end{table}

\begin{table}[ht!]
\begin{center}
\begin{tabular}{|lll|}
\hline
problem
& quantum complexity
& classical\\ \hline
breadth-first search 
& $O(\sqrt{VE\lg V})$
& $O(E)$ \\
depth-first search
& $O(\sqrt{VE\lg V})$
& $O(E)$ \\
single src.\ short.\ paths ($\pm$ wt.)
& $O(\sqrt{V^3E\lg V})$
& $O(VE)$ \\
all-pairs short.\ paths ($\pm$ wt.)
& $O(\sqrt{V^3E}\lg V + V^2\lg^3 V)$
& $O(V^{2.575})$ \\
\hline
\end{tabular}
\end{center}
\vskip-14pt
\caption{Graph theory in edge list model: change $E$ to $V^2$ for matrix model complexity}
\vskip10pt
\label{graph_table}
\end{table}

Note that several of our graph algorithms can run more slowly than their classical counterparts for $E$ sufficiently small; in each such case there is some $a$ such that the quantum algorithm is faster if $E\in\Omega(V\lg^aV)$.

\vspace{10pt}
\begin{table}[ht!]
\begin{center}
\begin{tabular}{|lll|}
\hline
problem
& quantum complexity
& classical\\ \hline
single src.\ short.\ paths ($+$ wt.)
& $O(\sqrt{VE}\lg V + V\lg^3V)$
& $O(E + V\lg V)$ \\
\ \ same, previous quantum
& \ \ $O(\sqrt{VE}\lg^2 V + ?)$
& \\
bipartite matching
& $O(V\sqrt{(E+V)\lg V})$
& $O((E+V)\sqrt{V})$ \\
\ \ same, previous quantum
& \ \ $O(V\sqrt{E+V}\lg V)$
& \\
\hline
\end{tabular}
\end{center}
\vskip-14pt
\caption{Improvements to quantum graph algorithms from other papers, in edge list model}
\vskip10pt
\label{graph_other_table}
\end{table}

\begin{table}[ht!]
\begin{center}
\begin{tabular}{|lll|} 
\hline
problem
& quantum complexity
& classical \\ \hline
points on a line ($\mathbb{Z}^n$)
& $N^{3/2}n\lg U$
& $N^2 n\lg U$ \\
points on a line ($\mathbb{R}^2$)
& $N^{3/2}\lg N$
& $N^2\lg N$\\
coin changer
& $D\sqrt{C\lg D}$
& $DC$ \\
maximum subarray sum
& $N^2$
& $N^3$ \\
\hline
\end{tabular}
\end{center}
\vskip-14pt
\caption{Computational geometry and dynamic programming}
\vskip10pt
\label{geom_table}
\label{dp_table}
\end{table}

In this paper we have chosen to focus on deriving new algorithms rather than proving lower bounds.  As such, it is possible that the algorithms presented here are not optimal, presenting clear directions for future research: searching for lower bounds that approach the upper-bounds presented here, and finding faster algorithms.  There are few published quantum algorithms (at least when viewed in the context of the number of published classical algorithms!) so there is limited sport to be had in picking them apart to save factors of $\sqrt{\lg N}$; on the other hand, there is a vast field full of classical algorithms with no quantum counterparts, and much of the low-hanging fruit remains untouched.

\section{Acknowledgements}

The author, being a neophyte to the art of writing scientific papers, would particularly like to thank his advisor, Bill Unruh, for being an excellent font of paper-writing advice and encouragement.  He would also like to thank Yury Kholondyrev, Matthew Chan and others involved in the UBC programming team for some early ideas of problems to tackle, and plenty of practice explaining himself.  Finally he would like to thank D\"urr and H\o yer, authors of \cite{dh}, for being the first to show him an exciting field, full of potential.

\begin{appendix}
\section{Detailed error analysis} \label{error}

Here we present in more exacting detail the parameters $\epsinv$ that are passed from function to function from section \ref{graph} and on, as well as brief (but complete) error analysis.  $\epsilon$ in this appendix will always denote the probability of failure for a function.  We pass the parameter $\epsinv$ rather than $\epsilon$ because $\epsinv$ is often polynomial in the input, and is thus more convenient to discuss.

\subsection{Graph algorithms} \label{error_graph}
Breadth-first search: in section \ref{bfs}'s step \ref{bfs_findall}, we call \findall.  It should be called with parameter $V\epsinv,$ giving the $V$ calls to it probability $1-\epsilon$ of all succeeding.  As this is our only function call that may fail, it gives the whole \bfs\ function probability $\epsilon$ of failure and running time $O\left(\sqrt{V^3\lg(V\epsinv)}\right)$ in the matrix model, $O\left(\sqrt{VE\lg(V\epsinv)}\right)$ in the edge list model.

Depth-first search: in section \ref{dfs}'s step \ref{dfs_findsol}, we call \findsol.  It should be called with parameter $2V\epsinv,$ giving the $2V$ calls to it (one to find each vertex, one from each vertex to find nothing) probability $1-\epsilon$ of all succeeding.  As this is our only function call that may fail, it gives the whole \dfs\ function probability $\epsilon$ of failure and running time $O\left(\sqrt{V^3\lg(V\epsinv)}\right)$ in the matrix model, $O\left(\sqrt{VE\lg(V\epsinv)}\right)$ in the edge list model.

Single-source shortest paths with negative edge weights: in section \ref{spnw}'s step \ref{spnw_assign}, we call \minfind.  It should be called with parameter $V^2\epsinv,$ giving the $V^2$ calls to it probability $1-\epsilon$ of all succeeding.  As this is our only function call that may fail, it gives the whole $SPNW$ function probability $\epsilon$ of failure and running time $O\left(\sqrt{V^5\lg(V\epsinv)}\right)$ in the matrix model, $O\left(\sqrt{V^3E\lg(V\epsinv)}\right)$ in the edge list model.

All-pairs shortest paths: in section \ref{apsp}, we call \spnw\ once and single-source shortest paths $V$ times.  Each should be called with parameter $(V+1)\epsinv,$ giving the $V+1$ total calls probability $1-\epsilon$ of all succeeding.  As these are our only function calls that may fail, they give the whole \apsp\ function probability $\epsilon$ of failure and running time $O\left(\sqrt{V^5}\left(\lg V + \sqrt{\lg(V\epsinv)}\right) + V^2\lg^3 V\right)$ in the matrix model, \\
$O\left(\sqrt{V^3E}\left(\lg V + \sqrt{\lg(V\epsinv)}\right) + V^2\lg^3 V\right)$ in the edge list model.

\subsection{Improvements to existing quantum graph algorithms} \label{error_graphother}

Single-source shortest paths: in section \ref{sssp}, we call \mindiff.  In D\"urr, Heiligman, H\o yer and Mhalla's notation\cite{dhhm}, for each size $s,$ we call \mindiff\ \;$n/s$ times; summing over sizes, we have $\sum_{k=0}^{\lg n}\frac{n}{2^k} < 2n$.  Switching back to our notation, that means we call it $2V$ times and require success each time, which means it should be called with parameter $2V\epsinv,$ giving the $V$ calls to it probability $1-\epsilon$ of succeeding.  As this is our only function call that may fail, it gives the whole function probability $\epsilon$ of failure and running time $O\left(\sqrt{VE}\left(\lg V + \sqrt{\lg(V\epsinv)}\right)+V\lg^3V\right)$.

Bipartite matching: in section \ref{bipartite}, we call \bfs\ and \dfs\ $\leq 2\sqrt{V}$ times each.  Each should be called with parameter $(4\sqrt{V})\epsinv,$ giving the $4\sqrt{V}$ total calls probability $1-\epsilon$ of all succeeding.  As these are our only function calls that may fail, they give the whole bipartite matching function probability $\epsilon$ of failure and running time $O(V\sqrt{(E+V)\lg(V\epsinv)})$.

\subsection{Computational geometry algorithms} \label{error_geometry}
Maximum points on a line in $\mathbb{Z}^n$: in section \ref{maxpoints_zn}, we call \minfind.  It should be called with parameter $\epsinv,$ giving the sole call to it probability $1-\epsilon$ of succeeding.  As this is out only function call that may fail, it gives the whole function probability $\epsilon$ of failure and running time $O(N^{3/2}n\lg U\sle)$.

Maximum points on a line in $\mathbb{R}^2$: in section \ref{maxpoints_r2}, we call \minfind.  It should be called with parameter $\epsinv,$ giving the sole call to it probability $1-\epsilon$ of succeeding.  As this is out only function call that may fail, it gives the whole function probability $\epsilon$ of failure and running time $O(N^{3/2}\lg N\sle)$.

\subsection{Dynamic Programming algorithms} \label{error_dp}
Coin changer: in section \ref{coinchange}'s step \ref{coinchange_minfind}, we call \minfind.  It should be called with parameter $D\epsinv,$ giving the $D$ calls to it probability $1-\epsilon$ of all succeeding.  As this is our only function call that may fail, it gives the whole \coinchange\ function probability $\epsilon$ of failure and running time $O(D\sqrt{C\lg(D\epsinv)})$.

Maximum subarray sum: in section \ref{subarray-sum}'s step \ref{subarray-sum_minfind}, we call \minfind.  It should be called with parameter $\epsinv,$ giving the sole call to it probability $1-\epsilon$ of succeeding.  As this is our only function call that may fail, it gives the whole \subarraysum\ function probability $\epsilon$ of failure and running time $O(N^2\sle)$.

\section{Comments and caveats} \label{caveats} \label{comments}
We mention here some comments that are important to the content of the paper, but that we felt broke up its flow too much to include in the body.

Asymptotic notation ($O,$ $\Omega$ and $\Theta$): informally, saying that a function takes $\Theta(f(N))$ time means that as $N$ goes to infinity, if we take the algorithm's running time and divide it by $f(N),$ we will get a nonzero constant; intuitively, that the function takes ``order'' $f(N)$ time to complete.  If a function takes $O(f(N))$ time, the algorithm's running time is upper-bounded by $f(N)$; $\Omega(f(N))$ is a lower-bound.  Throughout the paper we somewhat informally call our algorithms $O(f(N)),$ which we do because while the algorithm itself may be $\Theta(f(N)),$ the existence of that algorithm proves that the problem it solves is $O(f(N))$.  In section \ref{graph} we analyze many of our algorithms by saying they are better than the classical version for $E\in\Omega(V\lg^aV)$: this simply means that if we take the size of the graph to infinity, the algorithm is better as long as the number of edges goes to infinity at least as fast as $V\lg^aV$.

Types of graph: all graph algorithms presented here assume the graphs they operate on will have at most one edge between any two vertices (or two edges in opposite directions, in the directed case), and no ``self-edges'' $e_{aa}$.  Most of these algorithms are very easy to generalize to graphs that do not have that property, but in the interests of brevity we do not discuss that.

Large numbers: it is assumed throughout the body of the paper that basic arithmetic and addressing operations take constant time.  This is not the case as the size of input goes to infinity: take for example a graph with $2^{100}$ vertices.  Each vertex takes 100 qubits to address, and so looking at an edge out of $v_i$ is an $O(\lg V)$ operation.  The net effect of this is that every algorithm discussed in this paper has an unmentioned $\lg N$ (resp.\ $\lg V$) factor that we have not included in its running time.  In the literature when algorithms are analyzed, it is often the case that this extra factor is not included; so without opening that particular can of worms, we simply acknowledge that there is an extra factor of $\lg N$ everywhere, without putting it in the body of the paper.  Not including the extra factor is the tradition in much of classical computing, and is consistent with other papers on quantum algorithms (see for example \cite{dhhm,bbht,bcwz,as,grover,dh}).

\section{BBHT: probability of failure and running time} \label{bbht_error}

Here we explore, in some detail, the probability of failure and running time of Boyer, Brassard, H\o yer and Tapp's algorithm for quantum searching\cite{bbht}: in particular, their algorithm that finds one of an unknown number of solutions to a function $F$.  Recall that $F$ maps a domain of size $N$ to $\{0,1\},$ and has $M$ solutions $x$ such that $F(x)=1$; and that their algorithm runs in $O(\sqrt{N/M})$ calls to $F$.  The authors discuss average running time, but give scant attention to what happens if there is no solution; other papers (see for example \cite{bbw}) explore the algorithm in slightly more detail, but not to the degree we would like.  Here we attempt to encapsulate both average running time and probability of failure, as well as the running time's dependence on $\lambda,$ a constant chosen by the authors to be $8/7$.

In this appendix we assume familiarity with Grover's original algorithm\cite{grover}.  In particular, we ask that the reader be comfortable with the following:
\begin{itemize}
\item What it means to run Grover's algorithm with $j$ Grover iterations.
\item Let $\theta$ be such that $\sin^2\theta=M/N$.  Then the probability of success when Grover's algorithm is run with $j$ Grover iterations is $\sin^2((2j+1)\theta)$.
\end{itemize}

\subsection{The BBHT algorithm}
In the original algorithm, there is no provision for $M=0$; in that case, it runs forever.  We change this by inserting the condition $m>2\sqrt{N}$ (see below), at which point our algorithm decides there is no solution and returns {\it false}.
\begin{enumerate}
\item Initialize $m=1$ and set $\lambda=8/7$.
\item While $m \leq 2\sqrt{N},$ repeat the following unless instructed to return:
  \begin{enumerate}
  \item Choose an integer $j$ uniformly at random such that $0 \leq j < m$.
  \item Execute Grover's original algorithm, using $j$ Grover iterations.  Let the outcome be called $i$.
  \item If $F(i)=1,$ return $i$; otherwise, set $m$ to $\lambda m$.
  \end{enumerate}
\item Return {\it false}.
\end{enumerate}

Intuitively, \bbht\ works by trying several different numbers of Grover iterations, which (depending on how many iterations there were) will yield different probabilities of success for different values of $M$.  On average the algorithm as a whole will fail with probability $<.5M^{-.93},$ as we will see.

\subsection{Probability of failure and running time} \label{bbht_failure}

The probability of failure for \bbht\ is the probability that, for each $m$ up to $2\sqrt{N},$ Grover's algorithm never successfully returns a result when there is one to return.  To calculate that probability, first we need a result derived by Boyer, Brassard, H\o yer and Tapp\cite{bbht}: first, recall that after $j$ Grover iterations, the probability of returning a valid result is $\sin^2((2j+1)\theta)$.  For a given $m,$ $j$ could be any of $0\ldots m-1,$ and averaging over those values they arrive at a probability of $\frac12 + \frac{\sin(4m\theta)}{4m\sin(2\theta)}$ that an invalid result will be returned, for $m$ an integer.  $m$ is of course not actually an integer, but by choosing a random integer $0\leq j < m,$ we treat it as one and can consider it to be one for the purposes of that formula.

We wish to upper-bound the probability of error for \bbht\ as a whole, and we will start by differentiating between the cases $0 < \theta \leq \frac{\pi}4$ ($M\leq N/2$) and $\frac{\pi}4 < \theta \leq \frac{\pi}2$ ($M>N/2$).  For any $M\leq N/2,$ we wish to find an $m_0$ such that for each repetition of the outer loop when $m > m_0,$ the probability of failure is less than or equal to some constant.  For $M>N/2,$ we will find that the probability of failure is always less than or equal to some constant.

We begin by considering $M\leq N/2$.  In order to find $m_0,$ first we have to find critical points of $f_{\theta}(m)\equiv \frac12 + \frac{\sin(4m\theta)}{4m\sin(2\theta)},$ the probability that an invalid result will be returned:
\begin{align*}
\frac{df_{\theta}(m)}{dm}&=0\\
\frac{4\theta\cos(4m\theta)}{4m\sin(2\theta)} &= \frac{\sin(4m\theta)}{4m^2\sin(2\theta)}\\
4m\theta &= \tan(4m\theta)\\
4m\theta &= 0,4.49,7.73,\ldots
\end{align*}
Now we consider the form of $f_{\theta}(m)$.  It starts off at $f_{\theta}(0)=\frac12 + \frac{\theta}{\sin(2\theta)}$ and decreases from there; we want to find the first maximum it will return to after dipping down, meaning $4m\theta=7.73$.  Since $0 < \theta\leq\frac{\pi}4,$ we use $\sin(2\theta) \geq \frac{4}{\pi}\theta,$ and arrive at (when $4m\theta=7.73$) $f_{\theta}(m_0)\leq \frac12 + \frac{\sin(7.73)}{\frac{4}{\pi}\times 7.73} \approx 0.6$.  That does not give us $m_0,$ however: $m_0$ is when $f_{\theta}(m)$ first dips that low.  Solving numerically and using $\sin\theta\leq\theta$:
\begin{align*}
0.6 &= \frac12 + \frac{\sin(4m_0\theta)}{\frac{4}{\pi}4m_0\theta}\\
4m_0\theta &\leq 2.78\\
m_0 &\leq 0.69/\sin(\theta)\\
m_0 &\leq 0.69\sqrt{N/M}
\end{align*}
For $\frac{\pi}4 < \theta \leq \frac{\pi}2,$ although $f_{\theta}(m)$ is well-behaved and slowly-oscillating over the space of integer values of $m,$ it oscillates wildly in between; so our previous approach, based on considering $f_{\theta}$ as a function acting on the continuum, will not work.  To fix this problem, instead of considering $\theta,$ we now consider the angle $\phi\equiv\frac{\pi}2-\theta$; first noting that $f_{\theta}(m) = 1-f_{\phi}(m),$ meaning that success for $\theta$ corresponds to failure for $\phi$:
\begin{align*}
P_{\textit{fail}}(m) &= \frac12 + \frac{\sin(4m\theta)}{4m\sin(2\theta)} = \frac12 + \frac{\sin(4m(\frac{\pi}{2}-\phi))}{4m\sin(\pi-2\phi)} = \frac12 - \frac{\sin(4m\phi)}{4m\sin(2\phi)}
\end{align*}
Now we are back in the elysian realm of $0\leq\phi < \frac{\pi}2,$ and we can bound the probability of failure for $\phi$ from below and use that result.  The procedure here is as before, but instead of 7.73 we use the first root of $\tan(4m\phi)=4m\phi,$ 4.49.  For $\phi<\frac{\pi}4$ we use $\sin(2\phi) \leq 2\phi,$ and arrive at (when $4m\phi=4.49$) $P_{\textit{fail}}\geq \frac12 + \frac{\sin(4.49)}{2\times 4.49} \approx 0.39$.  That is the lowest the probability of failure $f_{\phi}(m)$ ever gets, and correspondingly it is the lowest the probability of success $1-f_{\theta}(m)$ ever gets.

We now have that, for any given iteration of the outer loop, the probability of failure for $M > N/2$ is less than or equal to 0.61 for all $m,$ and the probability of failure for $M \leq N/2$ is less than or equal to 0.6 for $m \geq m_0 = 0.69\sqrt{N/M}$.  We now compute the total probability of failure and running time for each case.

For $M>N/2,$ the total probability of failure is simply $0.61^{\lgl(2\sqrt{N})} \approx .5N^{\frac{-0.26}{\ln\lambda}},$ and the probability of getting to the $k\ith$ iteration through the main loop is $0.61^k$.  The total running time, then, is the sum $\sum_{k=0}^{\lgl(2\sqrt{N})} \frac{\lambda^k}{2}(0.61)^k < \frac12 \frac1{1-0.61\lambda}$.

For $M<N/2,$ the total probability of failure is $0.6^{\lgl(2\sqrt{N})-\lgl(0.69\sqrt{N/M})},$ which gives us $0.6^{\lgl(2.8\sqrt{M})} \approx (2.8M)^{-0.25/\ln\lambda}$.  The running time is the sum:
\begin{align*}
t&= \sum_{k=0}^{\lgl(0.69\sqrt{N/M})} \frac{\lambda^k}{2} + \sum_{k=\lgl(0.69\sqrt{N/M})}^{\lgl(2\sqrt{N})} \frac{\lambda^k}{2}(0.6)^{k-\lgl(0.69\sqrt{N/M})}\\
 &\approx \int_0^{\lgl(0.69\sqrt{N/M})} \frac{\lambda^k}{2}dk + \int_{\lgl(0.69\sqrt{N/M})}^{\lgl(2\sqrt{N})} \frac{\lambda^k}{2}(0.6)^{k-\lgl(0.69\sqrt{N/M})} dk\\
 &= \frac{0.69\sqrt{N/M}}{2\ln\lambda} + (0.69\sqrt{N/M})^{-\lgl 0.6} \int_{0.69\sqrt{N/M}}^{2\sqrt{N}} \frac{dx}{2} x^{\lgl 0.6}\\
 &= \frac{0.69\sqrt{N/M}}{2\ln\lambda} + (0.69\sqrt{N/M})^{-\lgl 0.6} \left[\frac{dx}{2}\frac{x^{1+\lgl 0.6}}{1+\lgl 0.6}\right]^{2\sqrt{N}}_{0.69\sqrt{N/M}}\\
 &= \frac{0.69\sqrt{N/M}}{2\ln\lambda} + \frac{(3\sqrt{M})^{\lgl 0.6}}{1+\lgl 0.6}\sqrt{N} - \frac12\frac{\sqrt{N/M}}{1+\lgl 0.6}
\end{align*}

Since we have $\sqrt{N/M}$ dependence from the first term, we should choose $\lambda$ such that the second term contributes no worse, which gives us the condition $\lgl 0.6 < 1,$ or $\lambda < 1.64$.  We now have:
\begin{align*}
t\leq\frac{0.69\sqrt{N/M}}{2\ln\lambda} - \frac12\frac{\sqrt{N/M}}{1+\lgl 0.6}
\end{align*}
which is minimal for $\lambda\approx1.31,$ and more importantly is $O(\sqrt{N/M})$.  We would like to note that this is about $50\%$ faster than Boyer, Brassard, H\o yer and Tapp's arbitrary choice of $\lambda=\frac87,$ but that is only true in this approximation; not only that, but the optimal value for $\lambda$ depends on the value of $M/N,$ so there is no one optimal $\lambda$ in general.

Using $\lambda=1.31,$ our results can be summarized in table \ref{bbht_result_table}.  Most important to us is that our running time is $O(\sqrt{N/M})$ calls to $F,$ and our probability of failure is less than $.5M^{-.93}$.  It is also worth noting that our earlier restriction, $\lambda < 1.64,$ came because we chose a small root for $\tan(x)=x$.  If we had chosen a larger root, $\lambda$ could have been larger, up to an asymptotic maximum of 2.

\begin{table}[ht!]
\begin{center}
\begin{tabular}{|ccc|} 
\hline
Case
& Probability of Failure
& Average Running Time \\ \hline
$M\leq N/2$
& $\leq .4M^{-.93}$
& $\leq 1.9\sqrt{N/M}$\\
$M>N/2$
& $\leq .5N^{-.96}$
& $\leq 2.3$\\
\hline
\end{tabular}
\caption{Probability of failure and average running time for BBHT, taking $\lambda$ to be 1.31}
\label{bbht_result_table}
\vskip10pt
\end{center}
\end{table}

\end{appendix}


\begin{thebibliography}{99}
\bibitem{kholondyrev}Thanks to Yury Kholondyrev for a discussion that led to this algorithm.
\bibitem{dustinmikematthew}Thanks to Wei-Lung Dustin Tseng, Michael Li and Man-Hon ``Matthew'' Chan for simultaneously suggesting that looking at Johnson's might yield something better than than the algorithm I was presenting to them at the time.
\bibitem{stevens}Thanks to Kory Stevens for a productive conversation that lopped off a factor of $\lg N$.
\end{thebibliography}
\end{document}